\documentclass[twocolumn,showpacs,preprintnumbers,prc,superscriptaddress]{revtex4}
\usepackage{epsfig}
\usepackage{graphicx}
\usepackage{epstopdf}
\usepackage{amsmath,amssymb,amsfonts}
\usepackage{array}
\usepackage{url}
\usepackage{hyperref}
\usepackage{csquotes}
\usepackage{xspace}
\usepackage{textcomp}
\usepackage{ulem}
\usepackage[usenames,dvipsnames]{color}
\usepackage{setspace}

\newcommand{\sqsn}{\mbox{$\sqrt{s_{_{NN}}}$}\xspace}
\newcommand{\bef}{\begin{figure}}
\newcommand{\eef}{\end{figure}}
\newcommand{\bc}{\begin{center}}
\newcommand{\ec}{\end{center}}

\newcommand{\auau}{\mbox{Au$+$Au}\xspace}
\newcommand{\Nudyn}{$\langle N_{\mathrm{ch}}\rangle \nu_{(\pm,\mathrm{dyn})}$\xspace}

\begin{document}
%\doublespacing
\title{ Fluctuations of identified particle yields using $\nu_{dyn}$ variable at energies available at the BNL Relativistic Heavy Ion Collider }
\author{Vivek Kumar Singh}
\email{vkr.singh@vecc.gov.in}
\affiliation{Variable Energy Cyclotron Centre, HBNI, 1/AF Bidhannagar, 
Kolkata 700 064, India}
\author{Dipak Kumar Mishra}
\email{dkmishra@barc.gov.in}
\affiliation{Nuclear Physics Division, Bhabha Atomic Research Center, 
Mumbai 400085, India}
\author{Zubayer Ahammed}
\email{za@vecc.gov.in }
\affiliation{Variable Energy Cyclotron Centre, HBNI, 1/AF Bidhannagar, 
Kolkata 700 064, India}

\begin{abstract}
We study the fluctuations of net charge, net pion, net kaon and net proton using 
the $\nu_{dyn}$ variable in the heavy-ion jet interaction generator (HIJING), 
ultra-relativistic quantum molecular dynamics (UrQMD), and hadron resonance gas (HRG) 
model at different collision energies \sqsn. It has been observed that the values of 
$\nu_{dyn}$ strongly depend on $\Delta \eta$ in the HIJING and UrQMD models and are
independent in the HRG model. The present work emphasizes the 
particle species dependence of net charge fluctuation strength and provides a 
baseline for comparison with the experimental data.

% \pacs{25.75.Gz,12.38.Mh,21.65.Qr,25.75.-q,25.75.Nq}
\end{abstract}
\maketitle
\section{Introduction}
\label{intro}  
One of the major goals of heavy-ion experiments is to study the phase transition 
from hadronic matter to  quark-gluon plasma (QGP). Event-by-event fluctuation of 
conserved quantities such as net-baryon number, net-electric charge, and 
net strangeness were proposed as  possible signals of the QCD phase 
transition~\cite{Jeon:2000wg}. Measurements of fluctuations can also help in understanding the nature of such a phase transition. One of the observables, net-charge fluctuation, 
has been considered as a signal for such studies. The reason behind net charge 
fluctuation study is similar to the original study of color charge in $e^{+}e^{-}$ 
experiment. There the color charge ratio was measured and depending upon the 
difference in fundamental 
degrees of freedom between quark-gluon state and hadronic state, the origin of the 
color charge was determined~\cite{Asakawa:2000wh,Jeon:1999gr}. Several experiments 
have measured net-charge  fluctuation at SPS, RHIC, and LHC 
energies~\cite{Adams:2003st,Abelev:2008jg,ex3,ex4,ex5,Abelev:2012pv}.

Event-by-event fluctuations in high-energy heavy-ion collisions have been used 
to study the equilibrium of thermodynamical fluctuations at freeze-out. In the QGP 
phase, quarks are the charge carriers with a fractional charge of $\pm 1/3$ or $\pm 2/3$ , while in the
hadronic phase hadrons are the charge carriers each with an integer 
charge. Hence, net-charge fluctuations in the QGP phase are predicted to be a factor 
of 2 to 3 smaller as compared to that of the hadronic phase 
~\cite{Jeon:2000wg}. These differences may be considered as 
indicators of the formation of quark-gluon plasma in high energy heavy-ion 
collisions. Thus, the net-charge fluctuations are strongly dependent on the phase of 
their origin. Due to the rapid expansion of the fireball created in the heavy ion 
collisions, the fluctuations created in the initial state may survive during the 
hadronization process~\cite{Asakawa:2000wh}. If the relaxation time happens to be 
shorter than the lifetime of the hadronic stage of the collisions, then the values of 
such fluctuations should deviate from their equilibrium hadron gas values towards 
their earlier, primordial values, typical for QGP~\cite{Asakawa:2000wh,Jeon:2000wg}. 
The fluctuations of different lengths or ranges in rapidity space relax on different 
time scales. Since relaxation can only proceed via diffusion of the charge, the 
longer range of fluctuations relaxes gradually. The relaxation time is expected to 
grow as a square of the rapidity range~\cite{Shuryak:2000pd}. It is evident that 
fluctuations of the total charge in a wider rapidity window relax slower. The minimal 
rapidity window that one can consider must be larger than the mean 
rapidity change of a charged particle in a collision, $\delta y_{coll}$. It is 
observed that the typical $\delta y_{coll}$ for the baryon and the electric charge is 
around 0.2 and 0.8, respectively~\cite{Shuryak:2000pd}. 

The Beam Energy Scan (BES) program at RHIC has been initiated to explore the QCD 
phase diagram and study the transport properties of nuclear matter at finite 
temperature ($T$) and baryonic chemical potential ($\mu_{B}$). At lower collision 
energy, e.g., \sqsn = 7.7 GeV, the baryon chemical potential can reach up to 
$\mu_B\sim$ 400 MeV, which is significant compared to the 
temperature of the fireball. At such energies, strong gradients in 
the chemical potential of conserved charges are expected. Hence, the lower beam 
energy scan program at RHIC will be useful to explore the properties of net-charge 
diffusion in nuclear matter. However, it is important to mention that the role of 
baryon stopping and long range correlations need to be explored extensively before 
making any conclusion on diffusion coefficients.

The conservation laws limit the dissipation of the fluctuations  after the 
hadronization has occurred. It is observed that, due to the diffusion of particles in 
rapidity space, these fluctuations may also get diluted in the expanding 
medium~\cite{Shuryak:2000pd,Aziz:2004qu}. The hadronic diffusion from the time of 
hadronization $\tau_0$ to a freeze-out time $\tau_f$ can dissipate these 
fluctuations.  It is argued that the reduction of the fluctuations in the QGP phase 
might be observed only if the fluctuations are measured over a large rapidity 
range~\cite{Shuryak:2000pd}. The work also quantifies the reduction of fluctuations  
with the increase of accepted rapidity interval. The suppression of charge 
fluctuations observed in the experimental data is consistent with the diffusion 
estimates~\cite{sup}. Earlier efforts were made to estimate the fluctuation  strength 
using a transport model for all inclusive charged particles~\cite{Mishra:2017bdq}. 
However, the contribution of different identified particles towards dilution of the 
measured fluctuation strength may be different.

One cannot measure the volume formed in heavy-ion 
collisions directly in the experiments. To avoid 
volume fluctuations, the ratio of positive ($+$) to negative ($-$) charged particles 
normalized by the total number of charged particles under consideration for a fixed 
centrality class of events is used to measure the fluctuation strength, usually 
known as $D$ measure~\cite{Jeon:2003gk}. This is defined as:
\begin{eqnarray}
D =  \langle N_{\mathrm{ch}}\rangle\langle \delta R^2\rangle &=& \frac{4}{\langle 
	N_{\mathrm{ch}}\rangle}\langle \delta N_+^2 + \delta N_-^2 - 2\delta N_+ \delta 
N_-\rangle \nonumber \\
&\approx& \frac{4\langle \delta Q^2\rangle}{\langle N_{\mathrm{ch}}\rangle},
\label{eq:Dmea}
\end{eqnarray}  

where $R (=N_+/N_-)$ is the ratio of number of positive particles to the number of 
negative particles, $Q = N_+ - N_-$ is the difference between the number 
of positive and negative particles (net charge) and $\langle N_{\mathrm{ch}}\rangle  
= \langle  N_+ + N_- \rangle $ is the average number of charged 
particles measured within the experimental acceptance. The $\langle \delta Q^2 
\rangle$ is the variance of the net charge $Q$, which is proportional to the 
net-charge fluctuation in the system. The value of $D$ is predicted to be 
approximately four times smaller in the QGP phase as compared to the hadron gas 
phase~\cite{Jeon:2003gk}. However, the $D$ measure has been found to be dependent on 
detection efficiency~\cite{Adams:2003st}. 

Another variable, $\nu_{(\pm,\mathrm{dyn})}$, is used to measure 
the fluctuation strength, which is robust and independent of detection efficiency. 
It is defined as 

\begin{equation}
\nu_{(\pm,\mathrm{dyn})} = \frac{\langle N_+(N_+ - 1)\rangle}{\langle 
	N_+\rangle^2} + \frac{\langle N_-(N_- - 1)\rangle}{\langle N_-\rangle^2} - 
2\frac{\langle N_-N_+\rangle}{\langle N_-\rangle \langle N_+ \rangle}.
\label{eq:nudyn}
\end{equation}
The value of  $\nu_{(\pm,\mathrm{dyn})}$ gives the measure of the relative 
correlation strength of (\enquote{$++$,} \enquote{$--$,} and \enquote{$+-$}) charged 
particle pairs. The relation between $D$ and $\nu_{(\pm,\mathrm{dyn})}$ is given as 
\cite{Jeon:2003gk} 

\begin{equation}
\langle N_{\mathrm{ch}}\rangle \nu_{(\pm,\mathrm{dyn})}\approx D - 4.
\end{equation}     
It is found that global charge conservation has a finite effect on the fluctuation 
variable $\nu_{(\pm,\mathrm{dyn})}$~\cite{Pruneau:2002yf, Bleicher:2000ek}. However, 
we have refrained from applying these corrections to our estimated values. One of the 
important aspects of this measured fluctuation strength is its survival 
probability. At high energy, i.e., in the limit $\langle N_+\rangle$ = $\langle 
N_-\rangle$, the magnitude of $\nu_{(\pm,\mathrm{dyn})}$ is determined by the 
integral of the balance function in the acceptance of the 
measurement~\cite{Pruneau:2002yf}. This integral depends on the relative width of the 
acceptance as well as the width of the balance function. The diffusion can further affect the value, 
but the magnitudes of $\nu_{(\pm,\mathrm{dyn})}$ are mainly 
determined by the 1/$N_\mathrm{ch}$ effect and charge conservation. Thus it is 
suggested to measure the fluctuation strength over large rapidity space 
which allows us to see deeper back into the history of the 
collision~\cite{Shuryak:2000pd}.

In the present study, we have calculated the fluctuation strength  of identified 
charged particles, mainly for net pion, net kaon and net proton using hadron 
resonance gas model (HRG), the heavy-ion jet interaction generator (HIJING) model, and the transport model ultrarelativistic quantum molecular dynamics (UrQMD) which will 
provide the reference for the behavior of fluctuations measured in the experiments.

The paper is organized as follows. In the following section, we discuss the HRG model 
used in this paper as well as the implementation of resonance decay. We also 
briefly discuss the HIJING and UrQMD models in the same section. In 
Sec.~\ref{sec:results}, we discuss our estimated results on 
$\nu_{(\pm,\mathrm{dyn})}$  for identified particles at different $\Delta\eta$ and 
\sqsn. We finally summarize our findings in Sec.~\ref{sec:summary}.

\section{ESTIMATION OF $\nu_{dyn}$ IN DIFFERENT MODELS}
\label{sec:diff models}
In this section, we briefly describe the models used in the 
calculation of $\nu_{dyn}$, which captures the strength of the correlations.  These 
models are extensively used to explain the experimental data from 
heavy-ion collisions.

\subsection{Hadron resonance gas model}
The partition function the in HRG model has all relevant degrees of freedom of the 
confined, strongly interacting matter and implicitly includes all the interactions 
that result in resonance formation~\cite{Karsch:2010ck,Garg:2013ata}. In the ambit 
of the grand canonical ensemble, the logarithm of the partition function is given as 

\begin{equation}
\label{eq:lnz}
\mathrm{ln} Z_i(T, V, \mu_i) = \pm \frac{Vg_i}{(2\pi)^3}\int d^3p 
~\mathrm{ln}\big\{1\pm \mathrm{exp}[(\mu_i-E)/T]\big\},
\end{equation}
where $i$ is the particle number index, $V$ is the volume of the system, $g_i$ 
is the degeneracy factor for the $i$th particle, $\pm ve$ signs correspond to the 
baryon or meson, respectively. We have used the total chemical potential of 
individual particle $\mu_i$ in our calculations as given in 
Ref.~\cite{Karsch:2010ck}. Using the partition function, one can calculate various 
thermodynamical quantities of the system in heavy-ion collisions. The 
susceptibilities of different orders are related to the $\langle N_\mathrm{ch} 
\rangle$ and $\langle \delta Q^{2} \rangle $ representing mean and variance of 
individual particle, respectively. These quantities can be calculated by taking the 
first and second derivative of Eq.(~\ref{eq:lnz}) with respect to $\mu$:

\begin{equation}
\label{eq:mean}
\langle N_\mathrm{ch} \rangle = \pm \frac{g_i}{2\pi^2}\int \frac{d^3p}
{\big\{1\pm \mathrm{exp}[(\mu_i-E)/T]\big\}},
\end{equation}
\begin{equation}
\label{eq:sig}
\langle \delta Q^2 \rangle = - \frac{g_i}{2\pi^2}\int \frac{d^3p}{T} 
\frac{\pm \mathrm{exp}[(\mu_i-E)/T]}{\big\{1\pm \mathrm{exp}[(\mu_i-E)/T]\big\}},
\end{equation}
Equations~(\ref{eq:mean}) and (\ref{eq:sig}) are used to calculate $\nu_{dyn}$ in HRG 
model.

Experimentally measured stable particles (pions, kaons and protons along with their 
anti-particles) have contributions from the production of both primordial as well as 
from resonance decay. Further, neutral resonances introduce positive correlations 
between $N_+$ and $N_-$ and hence, their decay daughters can affect the fluctuation 
of the final measured particles. The ensemble averaged stable particle yield will 
have contributions from both primordial production and the resonance 
decays~\cite{Begun:2006jf,Jeon:1999gr},
\begin{equation}
     \langle N_i\rangle = \langle N_i^*\rangle + \sum_R \langle N_R \rangle 
\langle n_i\rangle_R
\label{eq:ave}
\end{equation}
where $\langle N_i^*\rangle$ and $\langle N_R \rangle$ correspond to the average 
primordial yield of particle species $i$ and of the resonances $R$, respectively. 
The summation runs over all the resonances which decay to the final particle $i$ with 
$\langle n_i \rangle_R = \sum_r b_r^R n_{i,r}^R$ being the average number of particle 
type $i$ produced from the resonance $R$. Further, $b_r^R$ is the branching ratio of 
the $r$th decay channel of the resonance $R$ and $n_{i,r}^R$ is the number of 
particle $i$ produced in that decay branch. The generalized $n$th order 
susceptibility for stable particle $i$ can be written as~\cite{Mishra:2016qyj}

\begin{equation}
\chi_i^{(n)} = \chi_i^{*(n)} + \sum_R \chi_R^{(n)}\langle n_i\rangle ^n_R
\label{eq:chi_ave}
\end{equation}
The first term in Eq.(~\ref{eq:chi_ave}) corresponds to the contribution from 
primordial yield and the second term corresponds to the contribution from the 
fluctuation of primordial resonances and the average number of produced particle of 
type $i$, assuming the number of decay daughters is fixed.

\subsection{The HIJING and UrQMD models}
We have used HIJING (V.1.37) and UrQMD (V.1.30) to study the fluctuation variable 
$\nu_{dyn}$. Both HIJING and UrQMD models are Monte Carlo event generators used for 
nucleon-nucleon and nucleus-nucleus collisions in high energy physics simulations. 
These models provide a baseline to compare with the experimental data.

The HIJING model is based on perturbative QCD (pQCD) considering that the multiple 
mini-jet partons produced in collisions are transformed into string fragments and 
later, fragments into hadrons. It uses the PYTHIA model to generate kinetic 
variables for each hard scattering and the JETSET model for jet fragmentation. In pQCD, 
the cross section for hard parton scattering is determined using the leading order 
to account for the higher-order corrections. The soft contributions are determined 
using the diquark-quark strings with gluon kinks induced by soft gluon radiation. 
The HIJING model considers the nucleus-nucleus collisions as a superposition of 
nucleon-nucleon collisions; it also takes into account other physics 
processes like multiple scattering, jet quenching, and nuclear shadowing to study the 
nuclear effects~\cite{Wang:1991hta}.

The UrQMD model considers the microscopic transport of quarks and diquarks with 
mesonic and baryonic degrees of freedom. The model preserves the conservation of 
baryon number, electric charge, and strangeness number. In this model, the space-time 
evolution of the fireball is studied in terms of excitation and fragmentation of 
color strings, and the formation and decay of hadronic 
resonances~\cite{Bleicher:2000ek}. Interaction of the produced particles, which may 
influence the acceptance of certain windows, is included in the model. The formation 
of hadrons is explained by color string fragmentation, it also considers the 
resonance decays, multiple scattering between hadrons during the evolution including  
baryon stopping phenomena, which is one of the features of heavy-ion collisions 
especially at lower collision energies~\cite{Bleicher:1999xi}. The UrQMD model has 
been applied successfully to study the thermalization~\cite{Bravina:1998pi}, 
particle yields~\cite{Bass:1997xw,Soff:1999et}, leptonic and photonic 
probes~\cite{Spieles:1997ih}, and event-by-event 
fluctuations~\cite{Bleicher:1998wu,Xu:2016qjd,Zhou:2017jfk,Netrakanti:2014mta, 
Westfall:2014fwa,He:2017zpg}. 

It is noteworthy to mention that the measured values of fluctuation strength 
($\nu_{dyn}$) shall depend on the width of the acceptance, on the primordial 
mechanisms leading to $+ ve$ and $- ve$ particle production, radial transport (flow), 
diffusion, etc. HIJING and UrQMD models do not account for such effects explicitly. 

\section{Results and discussion}
\label{sec:results}
Due to the diffusion of charged particles in the hadronic phase, the measured 
fluctuation may get diluted during the evolution of the system and approaches
the equilibrated values in the hadronic medium until 
their kinetic freeze-out. Hence, the experimental measurements of the 
magnitudes of fluctuation strength at a fixed $\Delta\eta$ and 
their dependence on $\Delta\eta$ enable to us explore various aspects of the time 
evolution of the hot medium and the hadronization mechanism. It is proposed to study 
the fluctuations of identified particle species in different rapidity intervals.
%%%%%%%%%%%%%%%%%%%%%%%% Fig.1 %%%%%%%%%%%%%%%%%%%%%%%%%%%%%%
\bef[ht]
\bc
\includegraphics[width=0.5\textwidth]{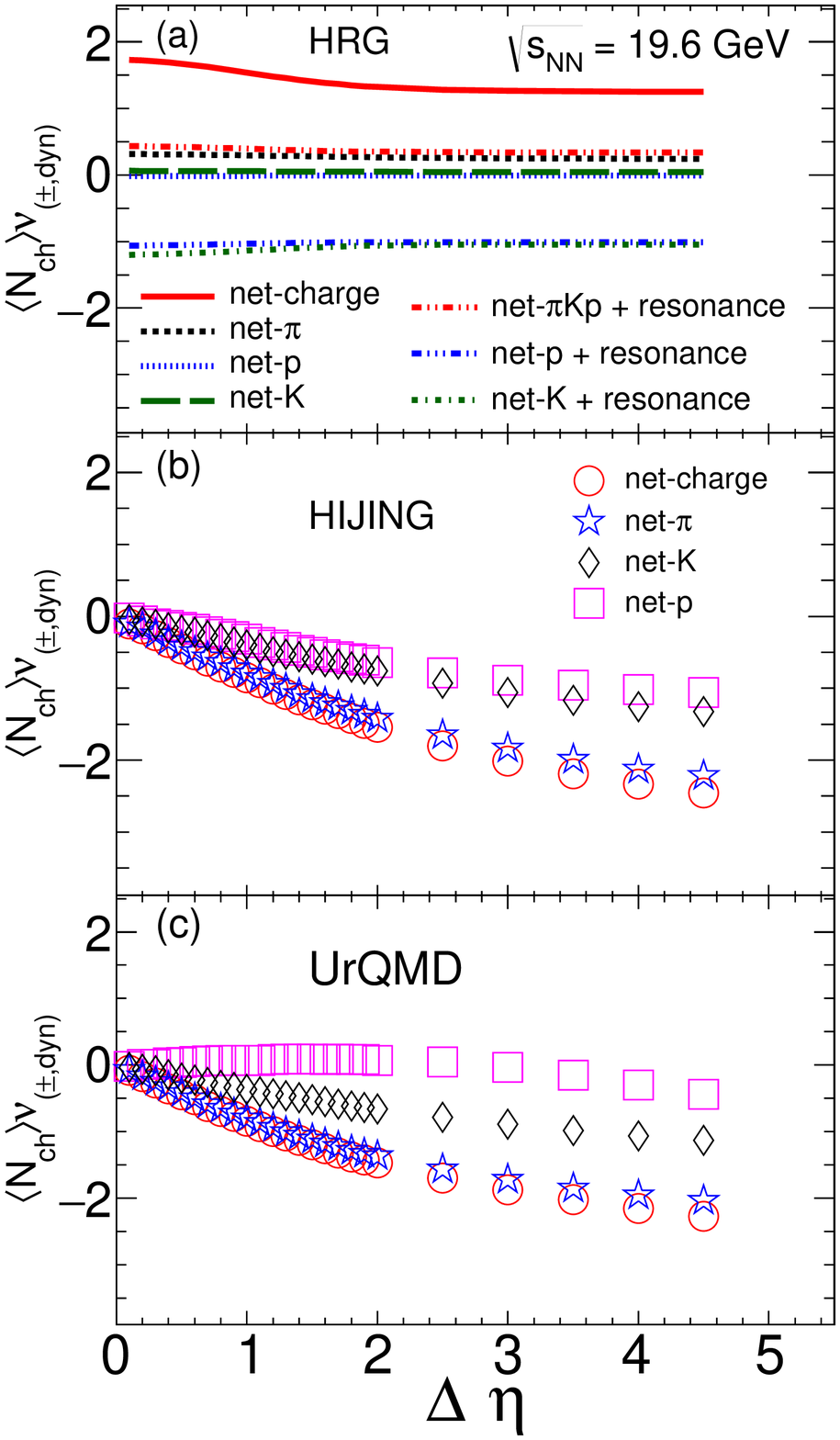}
\caption{ Fluctuation parameter \Nudyn as a function of $\Delta\eta$ for net-charge, 
net-pion, net-kaon and net-proton fluctuations for (0-5\%) centrality in Au$+$Au  
collisions at \sqsn = 19.6 GeV with HRG (upper panel), HIJING (middle panel) and 
UrQMD (lower panel) models. The \Nudyn from HRG model calculations 
for net-charge(solid line), net-$\pi$ and net-p(dotted lines), net-K(dashed line) without and  with resonance  
(dashed dotted lines) decay. The statistical errors are within symbol size.} 
\label{fig:NudynVsdeta19GeV}
\ec
\eef
%%%%%%%%%%%%%%%%%%%%%%%%%%%%%%%%%%%%%%%%%%%%%
%%%%%%%%%%%%%%%%%%%% Fig.2 %%%%%%%%%%%%%%%%%%%%%%%%%%%
\begin{figure}[ht]
\bc
\includegraphics[width=0.47\textwidth]{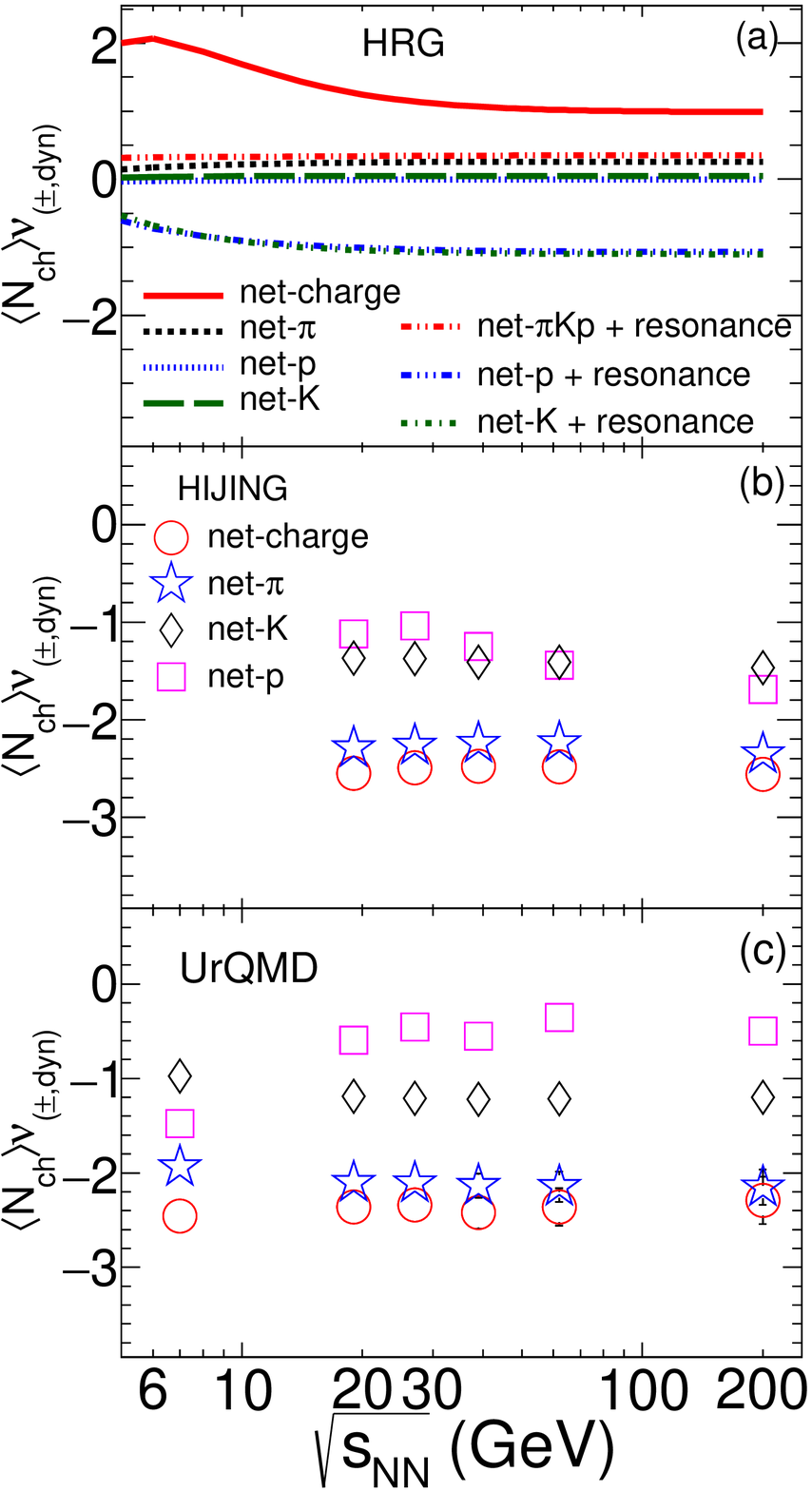}
\caption{Collision energy dependence of \Nudyn for net-charge, net-pion, net-kaon and 
net-proton are calculated using HRG (upper panel), HIJING (middle panel) and UrQMD 
(lower panel) models for (0-5\%) centrality in \auau collisions. The \Nudyn from HRG model calculations 
for net-charge(solid line), net-$\pi$ and  net-p(dotted lines), net-K(dashed line) without and  with resonance  
(dashed dotted lines) decay. The statistical errors are within symbol size.}
\label{fig:diff_Nudyn_ene}
\ec
\end{figure}
%%%%%%%%%%%%%%%%%%%%%%%%%%%%%%%%%%%%%%%%%%%%%%%%%%%%%
%%%%%%%%%%% Fig.3 %%%%%%%%%%%%%%%%%%%%%%
\begin{figure*}[ht]
\bc
\includegraphics[width=1.0\textwidth]{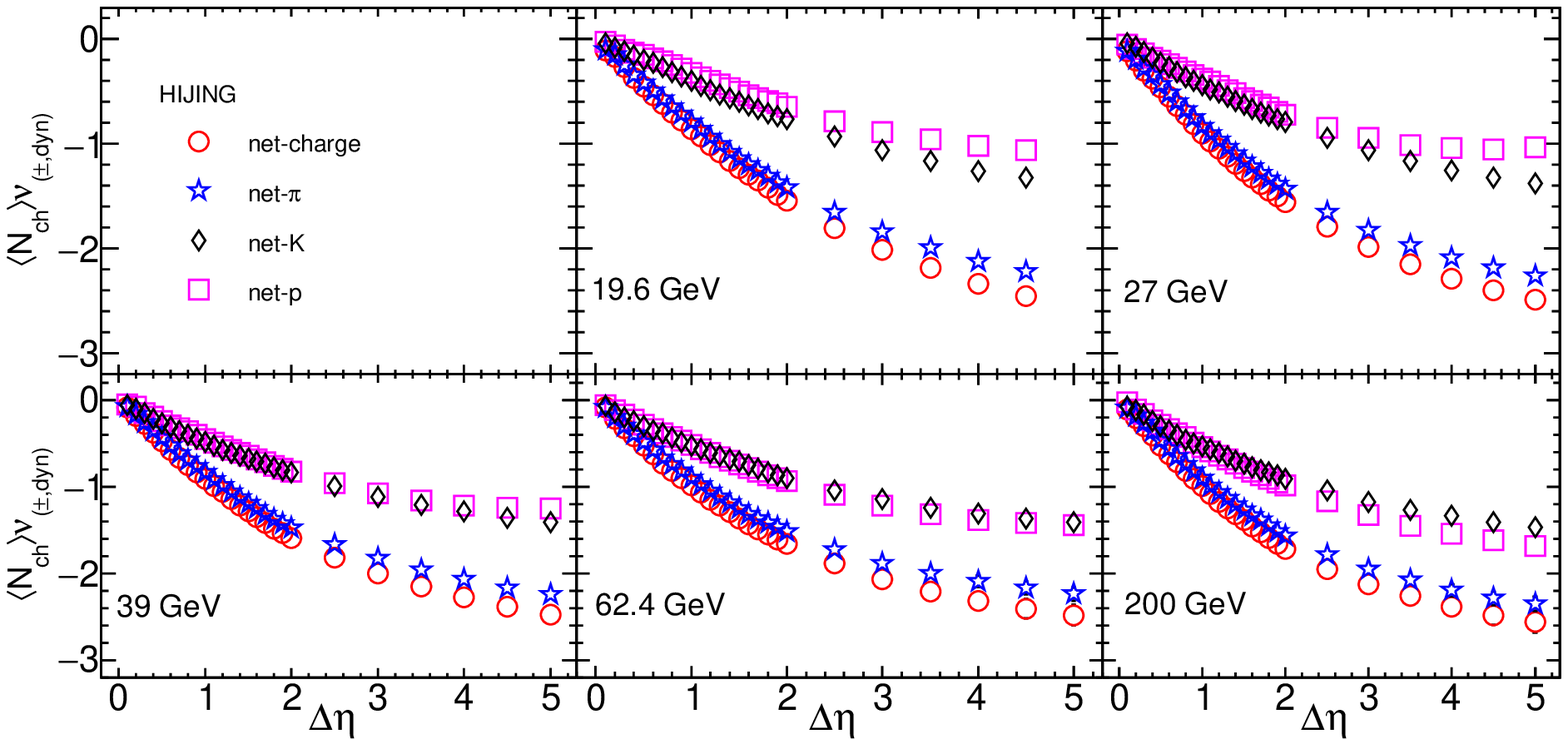}
\caption{The \Nudyn for net-charge, net-pion, net-kaon and net-proton as a function 
of $\Delta\eta$ window for (0-5\%) centrality in \auau collisions at different \sqsn 
in HIJING model. The statistical errors are within symbol size.}
\label{fig:NudynVsdeta@allE}
\ec
\end{figure*}
%%%%%%%%%%%%%%%%%%%%%%%%%%%%%%%%%%%%%%%%%%%%%
%%%%%%%%%%%%%%%%%%%%%% Fig.4 %%%%%%%%%%%%%%%%%%%
\begin{figure*}[ht]
\bc
\includegraphics[width=1.0\textwidth]{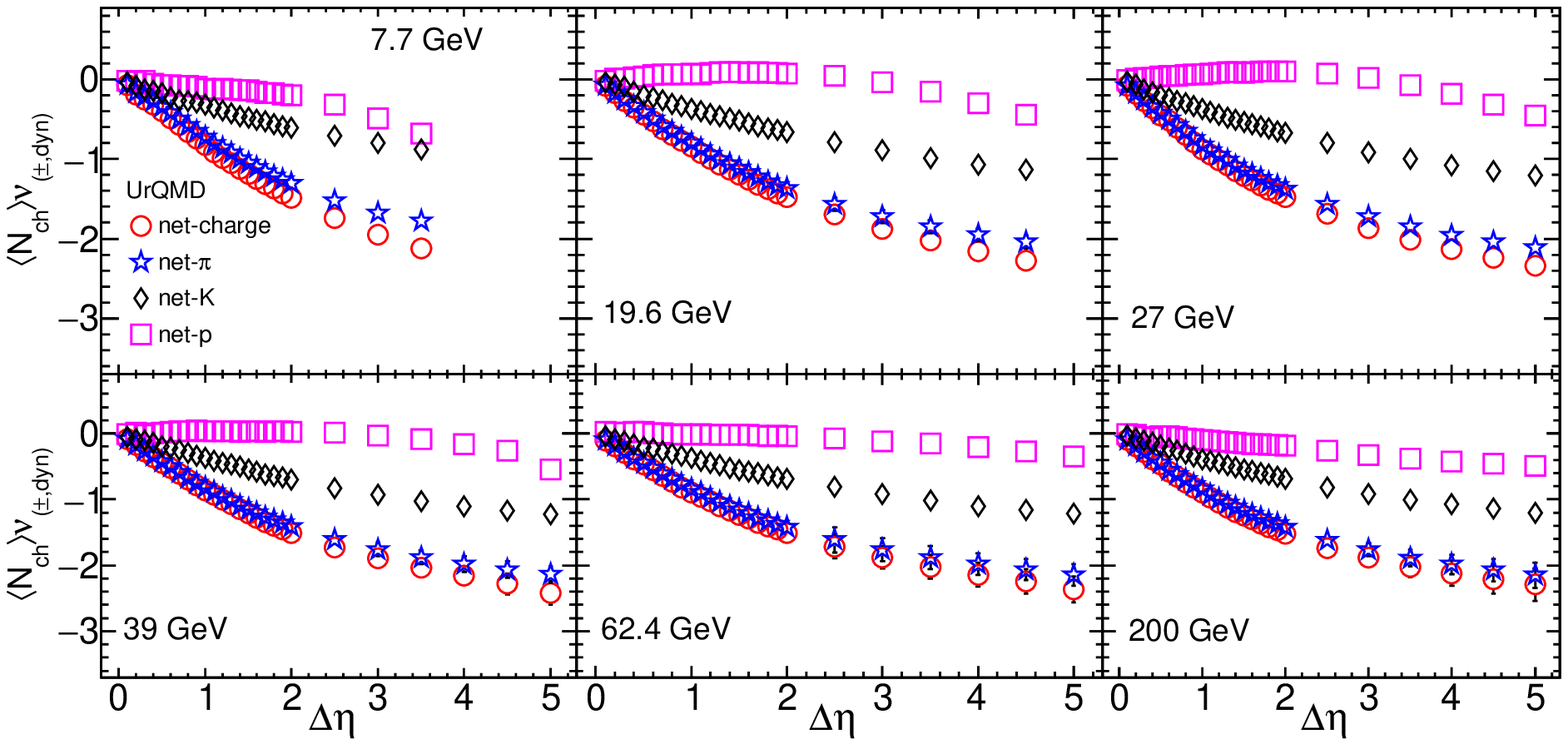}
\caption{The \Nudyn for net-charge, net-pion, net-kaon and net-proton as a function 
of $\Delta\eta$ window for (0-5\%) centrality in \auau collisions at different \sqsn 
in UrQMD model. The statistical errors are within symbol size.}
\label{fig:NudynVsdeta@allEUrQMD}
\ec
\end{figure*}
%%%%%%%%%%%%%%%%%%%%%%%%%%%%%%%%%%%%%%%%%%%%%%%%
%\subsection{$\Delta \eta$ dependence of fluctuation strength}

Figure~\ref{fig:NudynVsdeta19GeV} shows the  estimated value of  \Nudyn for 
net charge, net pion, net kaon and net proton as a function of $\Delta \eta$ 
interval with HRG, HIJING and UrQMD models at \sqsn = 19.6 GeV. 
For the present study, we have used 0.2 million central (0--5\%) \auau events 
at each energy in HIJING and UrQMD models. The particles having transverse momentum 
0.2 $\le p_T$ (GeV/$c$) $\le$ 5.0 are considered for the present study. The lower 
$p_T$ selection threshold is motivated by the existence of the experimental 
measurements performed at RHIC. The \Nudyn values estimated from HRG (upper panel), 
HIJING (middle panel), and UrQMD (lower panel) models are shown as a function of 
$\Delta\eta$. The HRG calculations for net-charge, net-pion, net-kaon, and net-proton 
fluctuations are performed within the same kinematic acceptance as those
used with HIJING and UrQMD models. Charged hadrons of masses up to 
2.5 GeV as listed in the particle data book are considered in the HRG model. The HRG 
model calculations are performed for different cases by considering all the charged 
hadrons, individual identified stable particles ($\pi , K, p$), and contribution of 
resonance decays to the stable particles. The estimated values of \Nudyn are found to 
be independent of $\Delta \eta$ in the case of the HRG model. However, there is a strong 
dependence of resonance decay effects observed for the identified particles. The 
calculation of \Nudyn from the HRG model will provide a pure thermal baseline 
contribution as a function of $\Delta \eta$. In the case of HIJING and UrQMD models, 
there is a strong dependence of \Nudyn values on $\Delta\eta$ are observed for 
net charge as well as for identified particles. The higher \Nudyn value at a 
lower $\Delta \eta$ interval suggests that the correlation is 
maximum for the smaller $\Delta \eta$ interval.

The curvature of \Nudyn shows a decreasing slope up to higher $\Delta \eta$ 
intervals. This is in contrast to the observation made by the ALICE 
experiment at higher collision energy \sqsn = 2.76 TeV,  which shows 
a flattening trend by extrapolating the fitted curve to higher $\Delta\eta$ 
range~\cite{Abelev:2012pv}. As can be seen, the \Nudyn values for net pion are closer 
to the results obtained for net-charge fluctuations. The net-charge fluctuation is 
dominated by the contribution from the pion fluctuation as the majority of the 
charged particles are pions. Similarly, the \Nudyn values for net kaon and net proton 
are closer to each other with a reduced slope as compared to net charge 
in the HIJING model. In the case of the UrQMD model, the slope of the \Nudyn 
values for net proton shows a flattening trend at a small $\Delta\eta$ and starts 
decreasing as a function of $\Delta\eta$ at a larger rapidity window. 

Figure~\ref{fig:diff_Nudyn_ene} shows the collision energy dependence of \Nudyn 
values for net charge and different identified net particles in most central \auau 
collisions using HRG (upper panel), HIJING (middle panel), and UrQMD (lower panel) 
models. The \Nudyn values for the net charge in HRG model decrease with increasing 
collision energies. In the case of identified particles (net $\pi$, net $p$ and net $K$) 
and contributions of resonance decay to these stable particles, the \Nudyn values do not change 
as a function of \sqsn. The \Nudyn values for net charge, net pion, and net kaon are 
independent of \sqsn in 
HIJING and UrQMD models. The \Nudyn values for the net-proton case show small energy 
dependence in both the models. There is a clear particle dependence of \Nudyn values 
for all collision energies in both the models.

Figures~\ref{fig:NudynVsdeta@allE} and \ref{fig:NudynVsdeta@allEUrQMD} show the 
\Nudyn as a function of $\Delta\eta$ intervals for (0-5\%) centrality in \auau 
collisions at different \sqsn using HIJING and UrQMD models, respectively. The 
$\Delta\eta$ dependence of \Nudyn for net proton is qualitatively different in both 
HIJING and UrQMD models, whereas net charge, net pion and net kaon 
show similar behavior in both the models. In the case of the UrQMD model, 
the \Nudyn values are flattened at higher $\Delta\eta$ with increasing \sqsn. For 
all the \sqsn, it is observed that the \Nudyn values of net charge and net pion 
have larger suppression as compared to net proton and net kaon. The observed 
suppression of \Nudyn for different particles may be due to the difference in the 
integral of the balance function of different identified 
particles~\cite{Pruneau:2002yf}.

\section{Summary}
% \vbox{
\label{sec:summary}
In summary, we have studied the fluctuations of net charge, net pion, net kaon, and 
net proton using the \Nudyn observable within the ambit of HRG, HIJING, and UrQMD 
models at different collision energies. The \Nudyn values are estimated up to a higher 
$\Delta\eta$ window. A stronger dependence of the \Nudyn value is observed for lower 
$\Delta\eta$ and the decreasing trend continues up to higher $\Delta\eta$ with the lower 
slope in both the models, except the net proton case in the UrQMD model. In the case of 
net proton in the UrQMD model, the curvature of \Nudyn values as a function of 
$\Delta\eta$ shows different behavior as observed in HIJING model. The \Nudyn values 
obtained from different model calculations are independent of collision energies but 
show particle species dependence. This study emphasizes the particle species 
dependence of  fluctuation strength and provides a reference baseline for comparison 
with the experimental data.

\begin{acknowledgements}
We are very much grateful to A. I. Sheikh and P. Garg for fruitful 
discussions and suggestions.
\end{acknowledgements}


\begin{thebibliography}{}

\bibitem{Jeon:2000wg} 
  S.~Jeon and V.~Koch,
  Phys.\ Rev.\ Lett.\  {\bf 85}, 2076 (2000).

\bibitem{Asakawa:2000wh} M.~Asakawa, U.~W.~Heinz and B.~Muller,
  Phys.\ Rev.\ Lett.\  {\bf 85}, 2072 (2000).
	
\bibitem{Jeon:1999gr} S.~Jeon and V.~Koch,
  Phys.\ Rev.\ Lett.\  {\bf 83}, 5435 (1999).

\bibitem{Adams:2003st} 
  J.~Adams {\it et al.} [STAR Collaboration],
  Phys.\ Rev.\ C {\bf 68}, 044905 (2003).

\bibitem{Abelev:2008jg} 
  B.~I.~Abelev {\it et al.} [STAR Collaboration],
  Phys.\ Rev.\ C {\bf 79}, 024906 (2009).

\bibitem{ex3} D. Adamove et al., Nucl .Phys. $\bf{A727}$, 97 (2003), H. Sako and H. Appelshäuser (CERES/NA45Collaboration), J. Phys. $ \textbf{G30} $, S1371(2004).
\bibitem{ex4} C. Alt et al., [NA49 Collaboration], Phys. Rev. $\bf{C70}$, 064903(2004).
\bibitem{ex5} K. Adcox et al., [PHENIX Collaboration], Phys. Rev. Lett. $\bf{89}$, 082301(2002),  K. Adcox et al.,[PHENIX Collaboration], Phys. Rev. $\bf{C66}$, 024901(2002).
   
\bibitem{Abelev:2012pv} 
  B.~Abelev {\it et al.} [ALICE Collaboration],
  Phys.\ Rev.\ Lett.\  {\bf 110}, 152301 (2013).

\bibitem{Shuryak:2000pd} 
  E.~V.~Shuryak and M.~A.~Stephanov,
  Phys.\ Rev.\ C {\bf 63}, 064903 (2001).

\bibitem{Aziz:2004qu} M.~A.~Aziz and S.~Gavin, Phys.\ Rev.\ C {\bf 70}, 034905 
(2004).
\bibitem{sup} Y.~Hatta and M.~A.~Stephanov, Phys.\ Rev.\ Lett.\  {\bf 91}, 102003 (2003), Erratum: [Phys.\ Rev.\ Lett.\  {\bf 91}, 129901 (2003)].
	
\bibitem{Mishra:2017bdq} 
         D.~K.~Mishra, P.~K.~Netrakanti and P.~Garg, Phys.\ Rev.\ C {\bf 95}, 054905 (2017).

\bibitem{Jeon:2003gk} S.~Jeon and V.~Koch, in \textit{Quark Gluon Plasma}, edited by R. C. Hwa and X. N. Wang (World Scientific, Singapore, 2004), pp. 430-490.
\bibitem{Pruneau:2002yf} C.~Pruneau,  Phys.\ Rev.\ C {\bf 100}, 034905 (2019);
 S. A. Bass, P. Danielewicz, and S. Pratt, Phys.\ Rev.\ Lett.\  {\bf 85}, 2689 (2000).
  
\bibitem{Bleicher:2000ek} M.~Bleicher, S.~Jeon and V.~Koch, Phys.\ Rev.\ C {\bf 62}, 061902 (2000).


\bibitem{Karsch:2010ck} F.~Karsch and K.~Redlich, Phys.\ Lett.\ B {\bf 695}, 136 (2011).
\bibitem{Garg:2013ata} 
  P.~Garg, D.~K.~Mishra, P.~K.~Netrakanti, B.~Mohanty, A.~K.~Mohanty, B.~K.~Singh and 
N.~Xu,
  Phys.\ Lett.\ B {\bf 726}, 691 (2013).
\bibitem{Begun:2006jf} 
  V.~V.~Begun, M.~I.~Gorenstein, M.~Hauer, V.~P.~Konchakovski and O.~S.~Zozulya,
  Phys.\ Rev.\ C {\bf 74}, 044903 (2006).
\bibitem{Mishra:2016qyj} 
  D.~K.~Mishra, P.~Garg, P.~K.~Netrakanti and A.~K.~Mohanty,
  Phys.\ Rev.\ C {\bf 94}, 014905 (2016).
	\bibitem{Wang:1991hta} X.~N.~Wang and M.~Gyulassy, Phys.\ Rev.\ D {\bf 44}, 3501 
(1991).
	\bibitem{Bleicher:1999xi} M.~Bleicher {\it et al.}, J.\ Phys.\ G {\bf 25}, 1859 (1999).
	\bibitem{Bravina:1998pi}  L.~V.~Bravina {\it et al.}, Phys.\ Lett.\ B {\bf 434}, 379 (1998).
	\bibitem{Bass:1997xw}  S.~A.~Bass {\it et al.}, Phys.\ Rev.\ Lett.\  {\bf 81}, 4092 (1998).
	\bibitem{Soff:1999et} S.~Soff, S.~A.~Bass, M.~Bleicher, L.~Bravina, M.~Gorenstein, E.~Zabrodin, 
	Phys.\ Lett.\ B {\bf 471}, 89 (1999).
	\bibitem{Spieles:1997ih}  C.~Spieles L. Gerland, N. Hammon, M. Bleicher, S. A. Bass,H. Stöcker, W. Greiner, C. Lourenço, and R. Vogt, Eur.\ Phys.\ J.\ C {\bf 5}, 349 (1998).
	\bibitem{Bleicher:1998wu} M.~Bleicher M. Belkacem, C. Ernst, H. Weber, L. Gerland, C.Spieles, S. A. Bass, H. Stöcker, and W. Greiner, Phys.\ Lett.\ B {\bf 435}, 9(1998).
	\bibitem{Xu:2016qjd}  J.~Xu, S.~Yu, F.~Liu and X.~Luo, Phys.\ Rev.\ C {\bf 94}, 024901 (2016).
	\bibitem{Zhou:2017jfk}  C.~Zhou, J.~Xu, X.~Luo and F.~Liu, Phys.\ Rev.\ C {\bf 96}, 014909 (2017).
	\bibitem{Netrakanti:2014mta} 
	P.~K.~Netrakanti, X.~F.~Luo, D.~K.~Mishra, B.~Mohanty, A.~Mohanty and N.~Xu,
	Nucl.\ Phys.\ A {\bf 947}, 248 (2016)
	\bibitem{Westfall:2014fwa} 
	G.~D.~Westfall,
	Phys.\ Rev.\ C {\bf 92}, 024902 (2015).
	\bibitem{He:2017zpg} 
	S.~He and X.~Luo,
	Phys.\ Lett.\ B {\bf 774}, 623 (2017).
	
\end{thebibliography}
\end{document}